\documentclass{appolb}
\usepackage{epsfig}

\begin{document}
\title{Combined analysis of $\eta$ and $\eta^\prime$ hadro- and photo-production off nucleons%
\thanks{Presented at the Symposium on Meson Physics, Cracow, 01-04 October 2008.}%
}
\author{K. Nakayama
\address{Department of Physics and Astronomy, University of Georgia \\ Athens, GA 30602, USA \\
Institut f\"ur Kernphysik, Forschungszentrum J\"ulich, 52425 J\"ulich, Germany}
\and
Yongseok Oh
\address{Cyclotron Institute and Physics Department, Texas {A\&M} University \\ College Station, TX 77843, USA}
\and
H. Haberzettl
\address{Center for Nuclear Studies, Department of Physics \\ The George Washington University,
Washington, DC 20052, USA}
}
\maketitle
\begin{abstract}
The production of $\eta$ and $\eta^\prime$ mesons in photon- and hadron-induced reactions are
studied in a combined analysis based on an effective Lagrangian approach in order to learn
about the relevant production mechanisms and the possible role of nucleon resonances in
these reactions.
\end{abstract}

\section{Introduction}

One of the primary motivations for studying the production of mesons off nucleons is to
investigate the structure and properties of nucleon resonances and, in the case of
heavy-meson productions, to learn about hadron dynamics at short range. In particular, a
clear understanding of the production mechanisms of mesons heavier than the pion still
requires further theoretical and experimental investigation. Apart from pion production,
the majority of theoretical investigations of meson-production processes are performed
within phenomenological meson-exchange approaches. Such approaches force us to correlate
as many independent processes as possible within a single model if one wishes to extract
meaningful physics information. Indeed, this is the basic motivation behind the
coupled-channel approaches.

Here, we present some selected results of our investigation of the $\eta$- and
$\eta^\prime$-meson productions in both the photon- and hadron-induced reactions,
\begin{eqnarray}
\gamma + N & \to & N + M  ,\nonumber  \\
\pi + N & \to  & N + M  , \label{eq:1}\\
N + N & \to  & N + N + M  \nonumber
\end{eqnarray}
based on a relativistic meson-exchange approach, where $M=\eta,\eta^\prime$.

We consider the three reactions mentioned above in the following manner:
the photoproduction reaction is calculated by considering the $s$-, $u$- and
$t$-channel Feynman diagrams plus a generalized contact term~\cite{NH1}, which
ensures gauge invariance of the total amplitude, in addition to accounting for
the final-state interaction (FSI) effects. (See Ref.~\cite{HNK} for details.)
The $\pi N \to N \eta$ reaction is described in the tree-level approximation
including the $s$-, $u$-, and $t$-channels. To the extent that this reaction
is dominated by the excitation of the $S_{11}(1535)$ resonance at least for
energies close to the threshold, this should be a reasonable approximation if
we confine ourselves to energies not too far from the threshold.
The $N N\to N N \eta$ process is calculated in the Distorted-Wave Born Approximation
(DWBA), where both the $NN$ FSI and the initial-state interaction (ISI) are
taken into account explicitly~\cite{NSL}. The $NN$ FSI is known to be responsible
for the dominant energy dependence observed in the total cross section (apart
from the dependence due to the phase space) arising from the very strong
interaction in the $S$-wave states at very low energies. As for the basic
meson-production amplitudes for the reactions in Eq.~(\ref{eq:1}),
our model~\cite{NH1,NOH1} includes the nucleonic, mesonic, and nucleon resonance currents
which are derived from the relevant effective Lagrangians.

The free parameters of our model --- the resonance parameters, the $NNM$
coupling constant, and the cutoff parameter $\Lambda^*_v$ at the electromagnetic
vector-meson exchange vertex --- are fixed such as to reproduce the available
data in a global fitting procedure of the reaction processes listed in
Eq.~(\ref{eq:1}).
The details of the present approach are fully described in
Refs.~\cite{NH1,NOH1}.

Ultimately, our goal is to perform a more complete model calculation where the
relevant FSIs are taken into account explicitly. However, before
undertaking a complex calculation that couples many channels, we need to learn
some of the basic features of meson productions within a simplified model where
these basic features
may be revealed and analyzed in an easier manner. In this regard, one of
the purposes of the present investigation is to show that consideration
of the hadronic reactions $NN \to NNM$, in conjunction with more basic two-body
reactions, would greatly help in the study of nucleon resonances, especially,
in imposing much stricter constraints on the extracted resonance-nucleon-meson
($RNM$) coupling strength involving a meson $M$ other than the pion. In fact,
our current knowledge of the branching ratios of the majority of the
known resonances is still very limited~\cite{PDG}.

\section{Results for $\eta$ meson production}

In this section we discuss some selected results for $\eta$-meson production
obtained according to the procedure described in the previous section.
We refer the details and the numerical results to Ref.~\cite{NOH1}.

\subsection{$\gamma p \to p \eta$}
For photoproduction of $\eta$ off nucleons, the available database is
considerable, in particular, for the proton target. This covers not only the
total and differential cross section data over a wide range of energy starting
from threshold but also the beam and target asymmetry data. When combined,
these data sets offer better opportunities for investigating the properties of
nucleon resonances. In particular, much more detailed studies than in the past
are possible now for resonances that may perhaps couple strongly to $N\eta$ but
only weakly to $N\pi$. In this respect, the recent data on the quasi-free
$\gamma n \to n \eta$ process~\cite{GRAAL} have attracted much interest in
$\eta$-production processes in connection to the possible existence of a narrow
(crypto-exotic) baryon resonance with a mass near $1.68$~GeV, which is still under
debate. To date, the narrow bump structure in the $\gamma n \to n \eta$ reaction
cross section was reported by several groups~\cite{LNS,CB-ELSA}.
In this work, we restrict our discussion to the $\gamma p \to p \eta$ reaction.

As far as the resonance currents are concerned, we follow the strategy adopted in
Ref.~\cite{NH1} to include the resonances one by one until a reasonable fit is achieved.
We found that the basic features of the total and differential cross section data
as well as the beam asymmetry data can be described by considering the $S_{11}(1535)$,
$S_{11}(1650)$, $D_{13}(1520)$, $D_{13}(1700)$, and $P_{13}(1720)$ resonances.
Inclusion of more resonances, especially, the higher spin resonances ($D_{15}$ and
$F_{15}$) did not change the overall quality of the fit. The spin-3/2 resonances are
important in describing the measured angular distributions in the range of $T_\gamma
= 1.07 \sim 1.6$ GeV and the beam asymmetry. But we found that the magnitude
of the measured beam asymmetry near threshold requires a more dedicated analysis.
Also, it is difficult to explain the measured target asymmetry; the same difficulty
has been encountered also in other investigations~\cite{Tiator99}. We emphasize that
the resonance parameters in our model are highly correlated to each other and
that the existing data are insufficient to establish a unique set of these parameters.
In Ref.~\cite{Tiator94}, cross section data at energies close to threshold have been
used to extract the $NN\eta$ coupling constant assuming that all the other relevant reaction
mechanisms are known at these energies. Close to threshold this reaction is
dominated by the $S_{11}(1535)$ resonance contribution.
Here, as an alternative, we show that the beam asymmetry at higher energies can also
constrain this coupling constant. The relatively small cross sections measured
at higher energies and backward angles favor a very small $u$-channel nucleonic
current contribution, compatible with a vanishing $NN\eta$ coupling constant.
However, we found that the cross section data at higher energies cannot constrain the
$NN\eta$ coupling constant uniquely because of a possible interference effect with the
nucleon resonance current.

\subsection{$\pi^- p \to n \eta$}

Most data for the $\pi N \to N \eta$ reaction have been obtained in the 1960s through
1970s; they are rather scarce and less accurate than for the $\gamma p \to p \eta$
and $NN \to NN \eta$ reactions. Recently, the Crystal Ball Collaboration has measured
the differential and total cross sections of this reaction near threshold~\cite{BNL_CB}.
Theoretically, this reaction has been studied mostly in conjunction with other
reactions in a combined analysis or in coupled-channel approaches (see, e.g., the references
quoted in Ref.~\cite{NOH1}) in order to constrain some model parameters. Recently, Zhong
\textit{et al.\/}~\cite{Zhong} have extended the chiral constituent quark model for meson
photoproduction to this reaction. Also, $\pi N \to N \eta$ has been studied within a
heavy-baryon chiral perturbation theory~\cite{Krippa}. Arndt \textit{et al.}~\cite{Arndt05}
have investigated the role of $\pi N \to N \eta$ on the $N\eta$ scattering length within a
coupled-channel analysis of this reaction and the elastic $\pi N$ scattering.

In the present work, the total cross section is nicely reproduced up to
$W\sim 1.6$~GeV, where it is dominated by the $S_{11}$ resonances, especially, the $S_{11}(1535)$.
In our study, the measured total cross section at higher energies are underestimated due to the absence
of the $\pi\pi N$ contribution via the coupled channel~\cite{Gasparyan}.
We also found that the $P_{13}(1720)$ resonance is important in reproducing the structure exhibited by the
measured angular distributions at higher energies.

\subsection{$N N \to N N \eta$}
This process is particularly relevant in connection to the role of the $\eta N$ FSI.
Most of the existing calculations take into account the effects of the $NN$ FSI in one way or
another, which is well-known to influence the energy dependence of the cross section near
threshold. Calculations which include the $\eta N$ FSI to the lowest order reproduce the bulk of the
energy dependence exhibited by the data. However, they fail to explain the $pp$ invariant mass
distribution measured by the COSY-TOF~\cite{COSYTOF} and COSY-11 Collaborations~\cite{COSY11}.
In Ref.~\cite{Fix}, the importance of the three-body nature of the final
state in the $S$-wave has been emphasized in order to account for the observed $pp$ invariant mass distribution.
Other authors have suggested an extra energy dependence in the basic production amplitude~\cite{Deloff}
to reproduce the existing data. Yet another possibility has been offered, which is based on a higher
partial wave ($P$-wave) contribution~\cite{NHHS}. We observe that all that
is required to reproduce the measured $pp$ invariant mass distribution is an extra $p^{\prime 2}$
dependence, where $p^\prime$ denotes the relative momentum of the two protons in the final state.
Obviously, this can be achieved either by an $S$-wave or by a $P$-wave contribution. Note that the
$NN$ $P$-wave $({}^3P_0$) can also yield a flat proton angular distribution as observed in the
corresponding data. As pointed out in Ref.~\cite{NHHS}, the measurement of the spin
correlation function should settle down the question on the $S$-wave versus $P$-wave contributions in
a model independent way.

Although the model calculation of Ref.~\cite{NHHS}, based on a stronger $P$-wave contribution, reproduces
nicely the shape of the measured $pp$ invariant mass distributions, it underestimates the total cross
section data near threshold ($Q < 30$~MeV). Here, we discuss the new results
based on a combined analysis of the $\gamma p \to p \eta$, $\pi^- p \to n \eta$, and $NN \to NN\eta$
reactions, which reproduce the currently existing data on $NN \to NN\eta$. (See Ref.~\cite{NOH1}.)
The major difference
from the previous calculation~\cite{NHHS} is a much stronger resonance contributions at low energies.
We emphasize that it is not possible to determine uniquely the resonance parameters involved in the present
work. In this process, in addition to the couplings of the nucleon resonances to pseudoscalar mesons, we need
their couplings to the vector mesons $\rho$ and $\omega$. As a consequence, we obtain either
the dominance of the $S_{11}$ resonance or $D_{13}$ resonance in the cross sections at low energies depending
on different interference patterns among the various meson exchanges in the excitation of these resonances.
The analyzing power is found to be very sensitive to these
interference patterns. Unfortunately, however, the existing data for this observable are not accurate
enough to disentangle them and more precise measurements of the analyzing power is strongly required.

\section{Results for $\eta^\prime$ meson production}

In contrast to the $\eta$ meson production, not many data exist for the $\eta^\prime$ meson
production reactions. In photoproduction, only the earlier total cross section data from ABBHHM
Collaboration~\cite{ABBHHM} were available until late 1990s. More recently, the differential cross
section data from the SAPHIR~\cite{SAPHIR} and the CLAS~\cite{CLAS} Collaborations became available.
Currently, the beam asymmetry in this reaction is being
measured by the CLAS Collaboration. Also, the CB-ELSA Collaboration has reported measurements of cross
sections in quasi-free $\eta^\prime$ photoproduction process on neutron~\cite{Krusche08}. For
$NN \to NN \eta^\prime$, there are data for total cross sections for excess energies up to $Q \approx 150$~MeV
and for the differential cross sections at two excess energies. Also, the $pp$ and $p \eta^\prime$
invariant mass distribution data will become available soon~\cite{Klaja}. But there are still no data for
$pn \to pn \eta^\prime$, apart from the very preliminary total cross section data by the COSY-11
Collaboration~\cite{Joanna}.

\subsection{$\gamma p \to p \eta^\prime$}
With the only data set available so far for $\gamma p \to p \eta^\prime$, it is not possible to have
stringent constraints on the resonance parameters. In fact, there are many sets of parameter values
which can reproduce the data equally well~\cite{NH1}. In order to impose more stringent constraints
on these parameters, one requires more exclusive and precise data, in particular, for the spin observables, such as
the beam asymmetry. A common feature of the model results corresponding to different sets of
parameters is the bump structure, which is likely to be due to the $D_{13}(2080)$ and/or $P_{11}(2100)$ resonance.
Therefore, experimental confirmation of this structure will be very interesting.
The angular distributions at higher energies and at large scattering angles restrict the value of the
$NN\eta^\prime$ coupling constant to be not much greater than $g_{NN\eta^\prime}^{} = 2$, which
is of particular interest in connection to the spin content of the nucleon.

\subsection{$NN \to NN\eta^\prime$}
In this reaction, the dominant production mechanism is found to be the excitation of the $S_{11}$
resonance. The energy dependence exhibited by the $pp \to pp\eta^\prime$ reaction is markedly
different from that of $pp \to pp\eta$. Both the (measured) total and differential cross
sections are rather well reproduced within our model. The preliminary data for $pp$ and $p\eta$ invariant mass
distributions~\cite{Klaja} are also rather well explained. It is interesting to note that,
within the statistical uncertainties, the $pp$ invariant mass distributions for $\eta$ and
$\eta^\prime$ productions exhibit similar features. The present model also reproduces the very
preliminary total cross section data (upper limit) for $pn \to pn\eta^\prime$.

\section{Results for vector meson production}

From the results for $NN\to NNM$ ($M=\eta,\eta^\prime$), it is clear that the present
investigation should be
extended to include the vector meson productions as well in order to impose more stringent
constraints on the resonance parameters. In particular, the $\omega$ and $\rho$ meson production
reactions will be useful to constrain the couplings of the nucleon resonances to these mesons
which are required to understand the $\eta$- and $\eta^\prime$-meson production mechanisms in $NN$
collisions. It is a part of our ongoing effort to consider pseudoscalar and vector meson productions
within a single model, and, here, we briefly report on the progress made in this direction so far.

The existing total and differential cross section data for the $\pi^- p \to n \omega$ reaction
can be described very well, consistently with the $\eta$-meson production reactions considered
in this work. However, the recent differential cross section data~\cite{SAPHIR} for $\gamma p
\to p \omega$ at higher energies and backward angles require a very small nucleonic current
contribution within the current approach, leading to a small $NN\omega$ coupling strength (or a very soft form factor~\cite{OTL})
which is
much smaller than those found in the description of other reactions such as $NN$ scatterings~\cite{Janssen}.
The present model does not consider the FSI; therefore, this could be a natural
explanation for the discrepancy. However, the calculations of Refs.~\cite{Giessen,Paris} based on
coupled-channel approaches, which take into account the FSI in one way or another, have also shown that
a much smaller $NN\omega$ coupling is needed in order to describe the photoproduction data.
Here we also
mention that the same problem is found in the description of $\rho$-meson photoproduction
within the present model. These findings indicate that this is not a trivial issue and that it
should be addressed in order to understand the mechanism of vector meson photoproduction.

\section{Concluding remarks}

A lot of data for meson production in photon-induced reactions have been accumulated and
many new data are expected
to be reported soon.
The existing world database for meson production (other than pion) from two-body \textit{hadronic\/} reactions, however,
is limited and most data are old and subject to rather large uncertainties. Currently there is a little effort to
improve and expand this database, which limits severely our ability to extract certain information
on nucleon resonances. In particular, information on the various decay rates (branching
ratios) of these resonances is strongly required. Meson productions in $NN$ collisions, therefore, can be used to improve this
situation: the existing data are rather accurate and the corresponding database can be expanded with
the existing facilities. So, this reaction should be included as a part of hadronic model calculations together with
photoproduction reactions
aiming at investigating nucleon resonances in conjunction with more basic two-body reactions.

\end{document}